\documentclass[conference]{IEEEtran}
\IEEEoverridecommandlockouts
\usepackage{cite}
\usepackage{amsmath,amssymb,amsfonts}
\usepackage{algorithmic}
\usepackage{graphicx}
\usepackage{subcaption}
\usepackage{textcomp}
\usepackage{xcolor}
\usepackage{comment}
\usepackage{tabularx}
\usepackage{xspace}
\usepackage{color, colortbl}

\usepackage{xcolor}
\begin{document}

\title{SCC: Automatic Classification of Code Snippets}

\newcommand{\dmg}[1]{{\color{red}#1}\xspace}

\author{Kamel Alreshedy, Dhanush Dharmaretnam, Daniel M. German, Venkatesh Srinivasan and T. Aaron Gulliver\\
University of Victoria\\
Victoria, BC, Canada V8W 2Y2\\
\{kamel, dhanushd, dmg, srinivas and agullive\}@uvic.ca}

\maketitle

\begin{abstract}
Determining the programming language of a source code file has been considered in the research community; it has been shown that Machine Learning (ML) and Natural Language Processing (NLP) algorithms can be effective in identifying the programming language of source code files. However, determining the programming language of a code snippet or a few lines of source code is still a challenging task. Online forums such as Stack Overflow and code repositories such as GitHub contain a large number of code snippets. In this paper, we describe Source Code Classification (SCC), a classifier that can identify the programming language of code snippets written in 21 different programming languages. A Multinomial Naive Bayes (MNB) classifier is employed which is trained using Stack Overflow posts. It is shown to achieve an accuracy of 75\% which is higher than that with Programming Languages Identification (PLI--a proprietary online classifier of snippets) whose accuracy is only 55.5\%. The average score for precision, recall and the F1 score with the proposed tool are 0.76, 0.75 and 0.75, respectively. In addition, it can distinguish between code snippets from a family of programming languages such as C, C++ and C\#, and can also identify the programming language version such as C\# 3.0, C\# 4.0 and C\# 5.0.


\end{abstract}

\begin{IEEEkeywords}
Classification, Machine Learning, Natural Language Processing, and Programming languages.
\newline

Tool demo video link: https://vimeo.com/275505431
\end{IEEEkeywords}

\section{Introduction}

In the last decade, Machine Learning (ML) and Natural Language Processing (NLP) based techniques have been widely applied in the study of source codes (cf. \cite{hindleNatural} \cite{Nguyen} \cite{probabilisticmodel} \cite{Learningtogenerate} \cite{Nguyen1}). It has been shown that these techniques can help with a variety of important tasks involving programming languages such as understanding the summary of source code, code completion and suggestion, code engine search and classifying programming languages.

Classifying programming languages of source code files using ML and NLP methods has been well explored in the research
community (cf. \cite{Kennedy}\cite{S.Gilda}\cite{Khasnabish}). It has been established that the programming language of
a source code file can be identified with high accuracy. However, most of the previous work that study the
classification of programming languages use the GitHub dataset in which the size of source code files is typically
large. Applying ML and NLP methods to classify a large source code file provides a very high accuracy as the large sample
contains many features that help the machine learning model to learn better. In this paper, we are interested in a tool that can classify a code snippet which is a small block reusable code with at least two lines of code, a much more challenging task. The only previous work that studies classification of the programming languages from a code snippet or a few lines of source code is the work of Baquero \textit{et al.}~\cite{Baquero}. However, they achieve low accuracy showing that identifying programming languages from a small source code or a code snippet is much harder than larger pieces. 

Predicting the programming language of code snippets accurately has several potential applications. For example, in
online social forums for programming languages such as Stack Overflow and Code Review, new users and novice developers
may forget to tag their posts. Predicting the programming language of code snippets inside the posts can  help predict a tag for the post.  
Tagging the Stack Overflow and Code Review questions with the correct 
programming language tag helps getting answers for a question.


Code snippet tools, such as gist and pastebin, allow users to  organize and share their code snippets with other users. These tools cannot predict the programming languages of these snippets and assume that the code snippets have already been tagged with the correct programming language by the user. 


Existing solutions to this pediction problem are not satisfactory. Integrated Development Environment (IDE) such as CLion,
Eclipse, and text editors such as Notepad++, SublimeText, Atom, predict the language based on file extension rather than
the source code itself. This can cause inconvenience to the users as they need to create the file with the correct extension manually to enable syntax highlighting in these editors.





The only known tool that can predict the programming language of a code snippet is Programming Languages Identification (PLI),
available in Algorithmia, a marketplace for AI based algorithms\cite{AlgorithmiaPLI}. PLI supports 21 programming languages: Bash, C, C\#, C++, CSS, Haskell, HTML, Java, JavaScript, Lua, Objective-C, Perl, PHP, Python, R, Ruby, Scala, SQL, Swift, VB, Markdown. It is claimed that PLI can predict 21 languages with a reported accuracy of 99.4\% top1 accuracy on GitHub source code. However, code snippets from Stack Overflow are much smaller in size compared to GitHub source code and the accuracy of PLI for code snippet classification has not been looked at. 



In this paper, we describe a new tool called Source Code Classification (SCC) to classify the programming language of a code snippet and compare our tool against PLI.  


The main contributions of this work are as follows:
\begin{enumerate}
\item
Description of a new tool, Source Code Classification (SCC), to classify the programming language of a code snippet from Stack Overflow. It uses a simple machine learning algorithm, Multinomial Naive Bayes (MNB), trained on Stack Overflow dataset and achieves an accuracy of 75\%, precision of 0.76, recall of 0.75 and F1 score of 0.75 in classifying $21$ programming languages. 
\item
Comparison of SCC to Programming Languages Identification (PLI) to show that SCC achieves much higher accuracy than PLI. PLI can only achieve an accuracy of 55.5\%, precision of 0.61, recall of 0.55 and F1 score of 0.55 in classifying $21$ programming languages. 
\item
Demonstrate that SCC can also distinguish between the family of programming languages, C, C\# and C++ with an accuracy of 80\%, and can identify the programming language version, C\# 3.0, C\# 4.0 and C\# 5.0 with an accuracy of 61\%.
\end{enumerate}

\section{Related Work}
Predicting a programming language from a given source code file has been a rising topic of interest in the research community.

Kennedy \textit{et al.} \cite{Kennedy} proposed a model to identify the software languages of entire source code files from Github using natural language identification techniques. Their classifier was based on five statistical language models from NLP trained on a GitHub dataset and identified $19$ programming languages with an accuracy of 97.5\%. S. Gilda \cite{S.Gilda} used a dataset from GitHub repositories for training a convolutional neural network classifier. Their classifier could classify 60 programming languages of source code files from Github with 97\% accuracy. 

J. N. Khasnabish \textit{et al.} \cite{Khasnabish}, collected more $20,000$ source code files to train and test their model. These source codes were extracted from multiple repositories in GitHub. The model was trained and tested using Bayesian classifier model and was able to predict $10$ programming languages with 93.48\% accuracy. 

D. Klein \textit{et al.}\cite{Klein} collected $41,000$ source code files from GitHub for the training dataset and $25$ source code files are randomly selected for testing dataset. However, their classifier, that used supervised learning and intelligent statistical feature selection, only achieved $48\%$ accuracy.

In \cite{Baquero}, the authors predicted the programming language of code snippets of Stack Overflow post. $1000$ question posts were extracted for each of $18$ different programming languages. They trained their classifier using a Support Vector Machine algorithm. Their model achieved very low accuracy of 44.6\% compared to the previous works because predicting the programming language of
a code snippet is more complex and challenging than predicting a source code file. 

\section{Tool Description}
\begin{figure}[!htb]
  \centering
  \frame{\includegraphics[width=9cm]{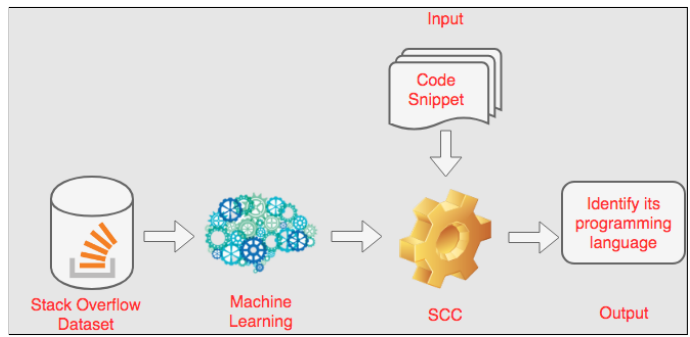}}
  \caption{Functioning of SCC. 
  }
  \label{fig:SCC}
\end{figure}

SCC is a classification tool to identify the programming language of code snippets. It is currently able to identify a code
snippet across 21 programming languages. SCC is an open source tool and therefore, it is possible to train it on a new dataset to support
and identify a new programming language. SCC is trained using a dataset curated from Stack Overflow and is implemented using Scikit-Learn \cite{Scikit-learn}, a machine learning library in Python.
Fig \ref{fig:SCC} shows how SCC functions. SCC is described in detail in the following subsections.

\begin{figure}[!htb]
  \centering
\frame{  \includegraphics[width=9cm]{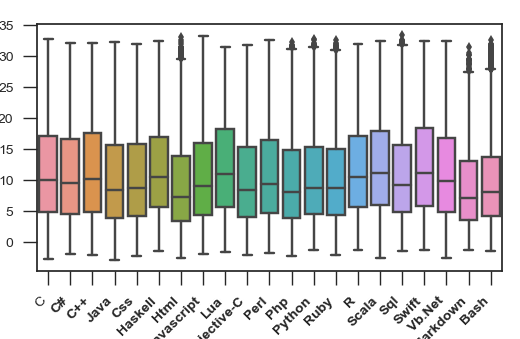}}
  \caption{The length of code snippets in the SO dataset. 
      }
  \label{fig:length}
\end{figure}

\subsection{Dataset}
We downloaded Stack Overflow data dump July 2017 `Post.xml'. Using Beautiful Soup library, 
we extracted $12000$ code snippets 
for each of $21$ programming languages from Stack Overflow posts; however, two languages, Lua (8428), Markdown (1359) had less than $12000$ posts. Question posts for each programming language were obtained using the question tag and by assuming the code snippets of Stack Overflow are properly tagged. For example, to get code snippets for Java programming, we searched for Java tag in Stack Overflow dataset. Question posts containing more than one programming language were removed to avoid any bias in our testing dataset. Fig \ref{fig:length} shows the average length of code snippet extracted for each programming language.

\subsection{Machine Learning}

Machine learning algorithms cannot learn from raw text; so several steps of processing the dataset are needed before training the algorithm. First, the code snippets need to be converted into numerical feature vectors. We used bag-of-word model, a method to represent each unique word as a feature. Second, we selected features extracted a subset from the dataset and used them for training the machine algorithm. The trained machine learning algorithm was in turn used to classify a new code snippet. The Scikit-Learn library \cite{Scikit-learn} was used to build SCC.

\textit{Multinomial Naive Bayes (MNB)} is a simple supervised machine learning algorithm based on Bayes theorem and commonly used in text classification and Natural Languages Processing (NLP). Each feature in a code snippet is assumed to be independent of the other features that occurs in the same snippet. It calculates the probability for each possible choice of programming language for a code snippet based on its feature vector and the programming language that has the maximum likelihood will be the final output. We chose MNB for its simplicity, speed and scalability properties.

Code snippets of Stack Overflow were split using the Term Frequency-Inverse Document Frequency (TF-IDF) vectorizer. The
most frequent ten words in each code snippet were selected. This helps machine learning algorithms to learn from the
most important words. The machine learning models were hypertuned using Grid-SearchCV - a tool for parameter search in
Scikit-learn. For a Multinomial Naive Bayes classifier, it is important to tune a hyper parameter called alpha (Additive
smoothing) parameter. These parameters were fixed after performing GridSearch on the Cross validation sets (10 fold
cross validation).

\subsection{Usage Example}
\begin{figure}[!htb]
  \centering
  \frame{\includegraphics[width=9cm]{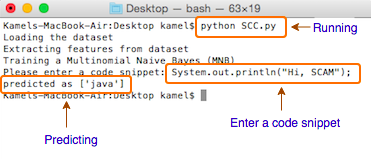}}
  \caption{How SCC works.
  }
  \label{fig:SCCDemo}
\end{figure}

SCC is a simple command-line tool, and was built based on Stack Overflow dataset and Multinomial Naive Bayes classifier. To run SCC, the first step is to load the dataset and select the feature set. The next step is to train the machine learning algorithm on the selected features. Subsequently, users will be asked to enter their code snippet through command line. Finally, the predicted programming language for the snippet is output. Fig \ref{fig:SCCDemo} demonstrates how SCC works. SCC is an open source\footnote{https://github.com/Kamel773/SourceCodeClassification} and dataset is available online\footnote{https://drive.google.com/open?id=1leMs0rdKAfX1UYEhSEe9a1npIWrIvqr6}.


\definecolor{Gray}{gray}{0.9}
\definecolor{LightCyan}{rgb}{0.88,1,1}

\section{Evaluation}


\subsection{Evaluation of SCC}


\subsubsection{Methodology}

SCC was evaluated to answer three research questions. For the first research question, our goal was to evaluate if SCC can classify a code snippet across $21$ programming languages. The purpose of the second research question was to evaluate if SCC is able to distinguish between a code snippet across one family of programming languages, C, C\# and C++. To answer this research question, we created a subset of our dataset which only contains three programming languages C, C\# and C++. Then, the dataset was split into training and testing datasets into ratio of 80:20. The final research question was to evaluate if SCC can identify different versions of one programming language, C\# 3.0, C\# 4.0 and C\# 5.0. A new dataset was extracted from Stack Overflow using three question tags, C\# 3.0, C\# 4.0 and C\# 5.0, to answer the third research question. 

\subsubsection{Results}
\ \\
\indent\textit{Can SCC identify the programming language of a code snippet?}

\begin{figure*}[!htb]
  \centering
 \frame{ \includegraphics[width=\textwidth]{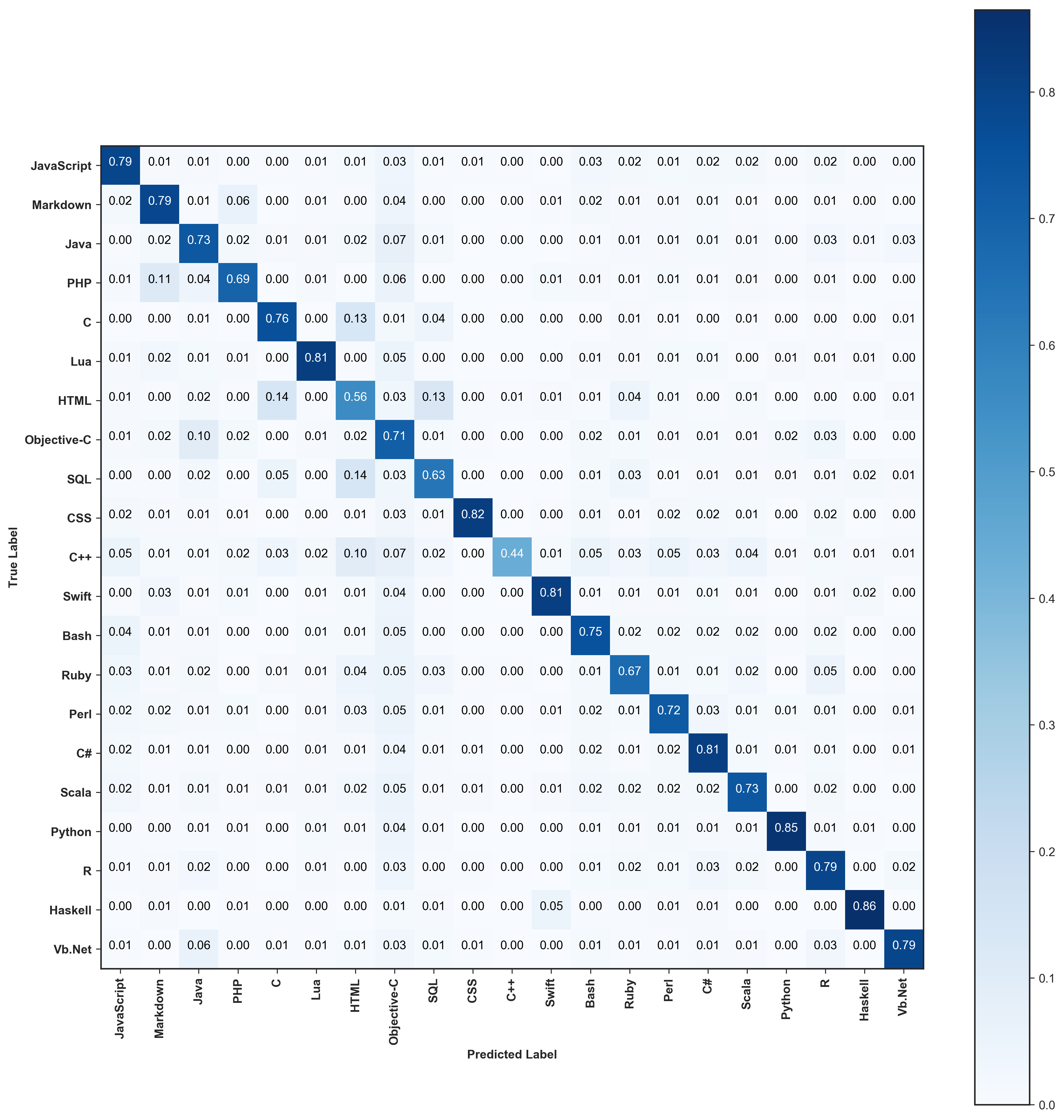}}
  \caption{Confusion matrix for all $21$ programming languages }
  \label{fig:Text}
\end{figure*}

To evaluate SCC, the dataset was split into training and testing data of the ratio of 80:20, that is, 20\% of code
snippets from each programming language was used for evaluation purpose. The resulting confusion matrix is shown in Fig \ref{fig:Text} and the scores obtained for precision, recall and F1-score are shown in Table \ref{my-label}. Overall, SCC achieved an accuracy of 72.0\% and the average scores for precision, recall and F1-score were 0.73, 0.72 and 0.72 respectively.

Several programming languages were classified with very high F1-scores - Haskell (0.89), Python (0.88), CSS (0.86),  Lua (0.86) and Swift (0.84). 

Objective-C had a much higher recall than precision, and these were 0.48 and 0.71 respectively. This is because 10.0\% of code snippets from Objective-C were misclassified as PHP. Furthermore, many code snippets from other programming languages were misclassified as Objective-C. When we examined these snippets, we noticed that either they were very small in size or their feature sets were common to many languages. Whenever this occurred, SCC misclassified them as Objective-C. The worst F1-score of 0.51 was for C++. For C++, precision score of 0.63 was much higher than recall score of 0.44. Many code snippets for C++ were classified incorrectly as JavaScript, Objective-C, Bash and Perl. SCC misclassified code snippets from HTML as SQL and C with percentages 13\% and 14\%. Also, 14\% of code snippets from SQL and 13\% from C were misclassified as HTML. 

\textit{Can SCC distinguish between code snippets across one family of programming languages such as C, C\# and C++?}

\begin{figure}[!htb]
  \centering
  \frame{\includegraphics[width=7cm]{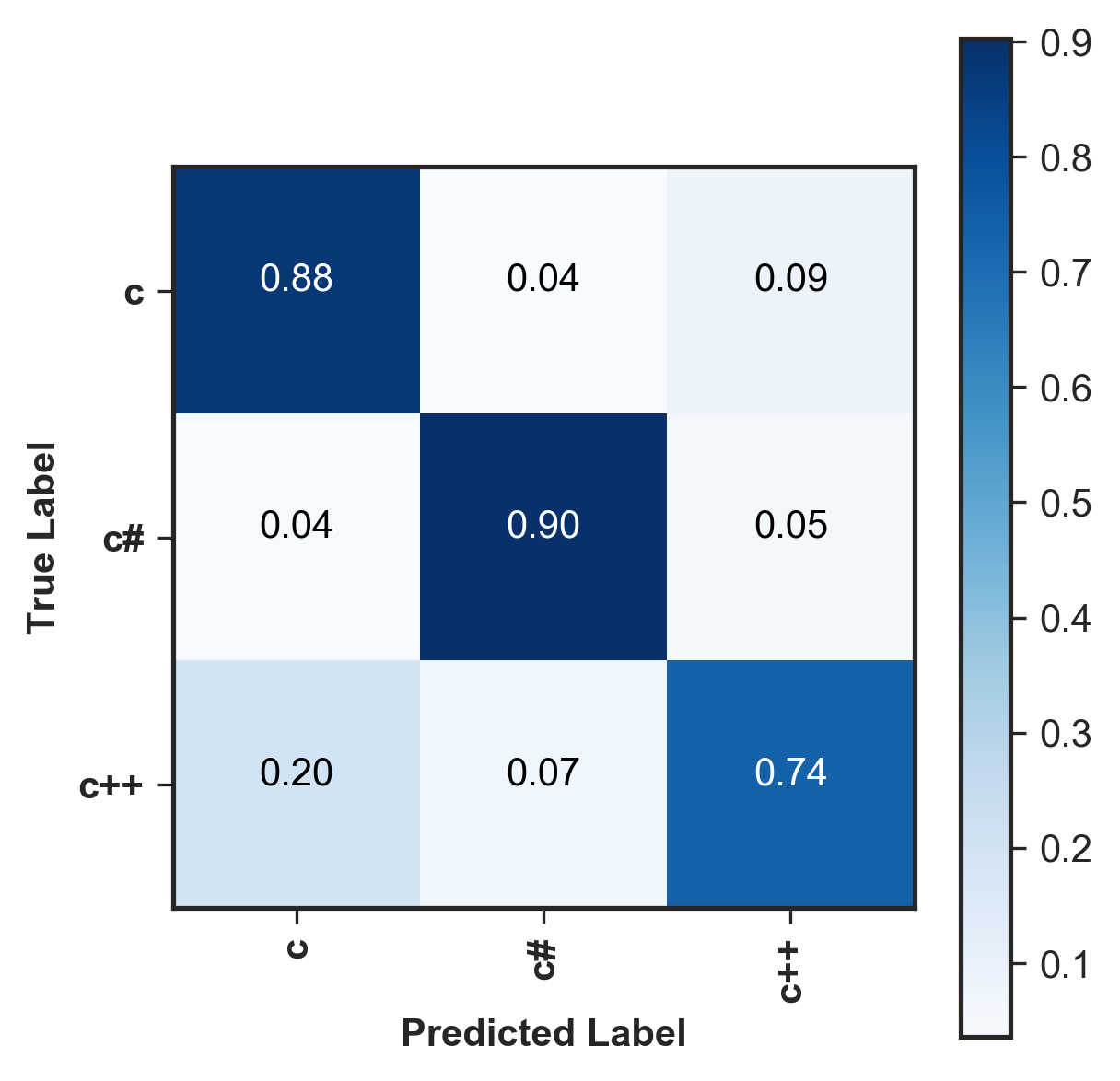}}
  \caption{Confusion matrix for C, C\# and C++}
  \label{family:cfu}
\end{figure}

This experiment involved training and testing SCC on three programming languages from the same family, C, C++ and
C\#. In this experiment SCC achieved a high accuracy of 80.0\% and the average scores for precision, recall and F1-score were 0.81, 0.80 and 0.80 respectively. Table \ref{family:pre} 
shows the details of the performance on C, C\# and C++. Also, the confusion matrix is shown in Figure \ref{family:cfu}. C\# has unique features compared to C and C++ and hence SCC can classify C\# with a high F1-score of 0.88. The percentage of C++ code snippets misclassified as C was 20\%. Table \ref{table:features1} 
shows the top $10$ features for these programming languages; these features helps SCC to learn and distinguish between these programming languages.

\textit{Can SCC identify among different versions of a programming language specifically C\#?}

\begin{figure}[!htb]
  \centering
  \frame{ \includegraphics[width=7cm]{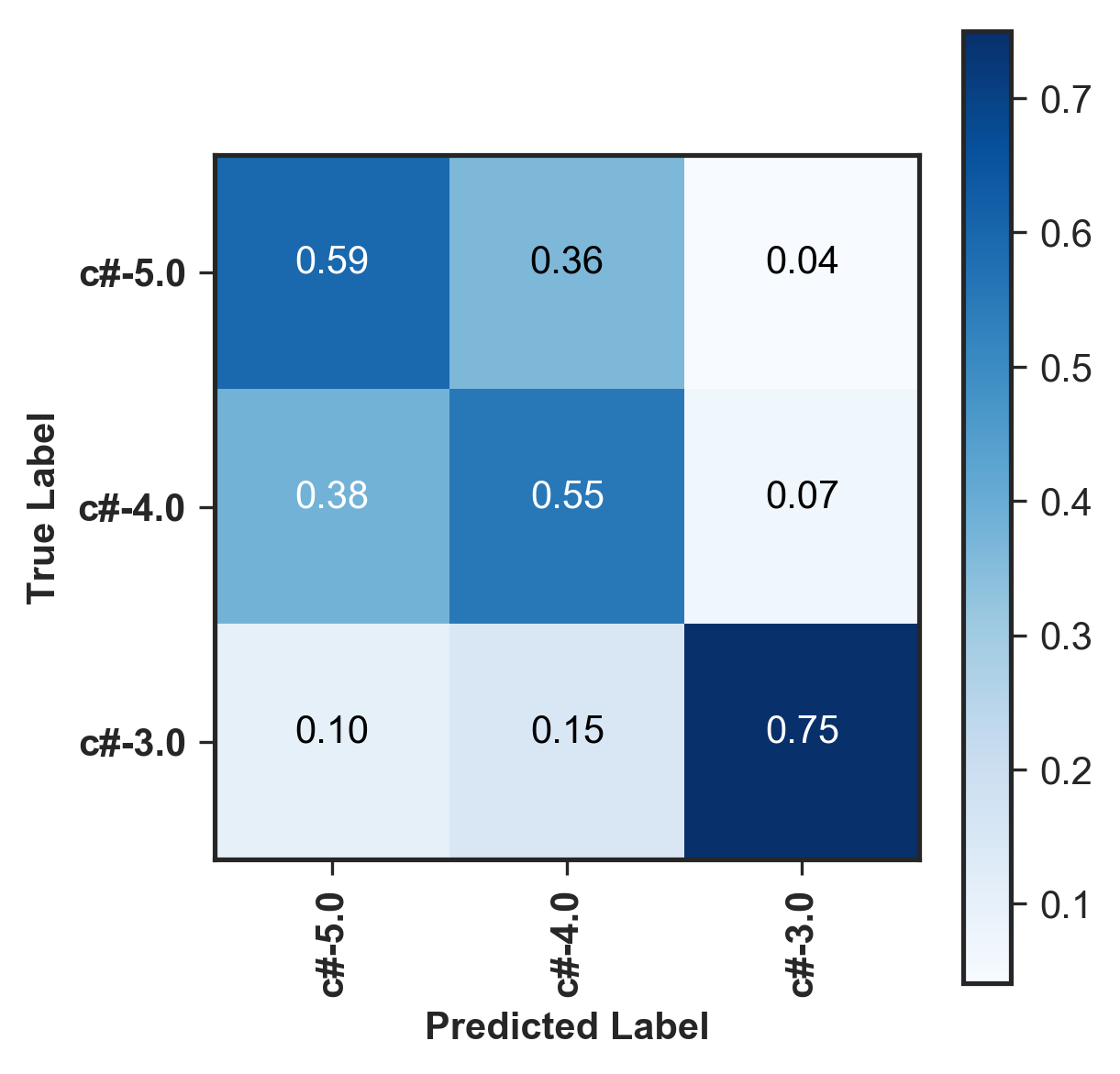}}
  \caption{Confusion matrix for C\#3.0, C\#4.0 and C\#5.0}
  \label{fig:Text1}
\end{figure}

For this experiment, SCC was trained and tested on a dataset that only contain three versions of C\#. SCC achieved an
accuracy of 61.0\% and the average of scores for precision, recall and F1-score were 0.61, 0.61 and 0.61
respectively. The confusion matrix is shown in Fig \ref{fig:Text1} and the details of the performance is shown in Table
\ref{cvarious}. We noticed that SCC finds it particularly hard to classify between the versions, C\# 4.0 and C\# 5.0. These results show that, while SCC is highly accurate in distinguishing between code snippets from C, C\# and C++ family, it is less accurate at identifying the different versions, C\# 3.0, C\# 4.0 and C\# 5.0. 


\subsection{Comparison to PLI}

\subsubsection{Methodology}

As mentioned in the introduction, it is claimed that PLI provides a high accuracy while predicting the programming languages from a give source code file. However, predicting the language of a code snippet is far more challenging. 
We evaluated the performance of PLI in predicting the programming languages of code snippets and compared its results
with SCC. $150$ code snippets were randomly selected from each programming language. This created a subset of $4200$
code snippets which could be used for prediction using PLI. We used only $150$ snippets because we had to pay for the
classification of each one of them. We used the urlib library in python to make the API calls to PLI to generate
predictions for all $4200$ code snippets. The API call returned a JSON file with languages as key and corresponding
probability for all $21$ languages that it supported. The programming language with highest probability score was
selected as the predicted language.  We could compare SCC with PLI with respect to the first research question because
both tools support the $21$ programming languages. However, we could not study the performance of PLI regarding the
second and third research questions because this tool is closed-source and we were not able to train it using
a dataset that contained only a family of programming language or different versions of a programming language.

\subsubsection{Results}
PLI achieved an accuracy of 55.5\% and the average scores for precision, recall and F1-score were 0.61, 0.55 and 0.55 respectively. 
The performance of PLI for each programming language in shown in Table~\ref{my-label}.  


CSS had the worst recall score of 0.19 among the programming languages. This is because 40\% of code snippets of CSS were misclassified as HTML. The syntax and operations of these two programming languages are extremely similar to each other. Similarly, 21\% of the code snippets of JavaScript were misclassified as HTML. We also noticed that PLI misclassified many code snippets from CSS and HTML as Javascript, pointing to its inherent weakness.

Objective-C had the highest F1-score of 0.77. This language has very unique syntax compared to other
programming languages in our study. 
Some programming languages that were correctly classified with high precision were Vb.net (0.89), R (0.88), Objective-C (0.85) and Bash (0.79).


19\% of code snippets from C were classified as C++, and 7\% of code snippets of C++ were classified as C. Ruby, HTML, CSS and Markdown had the worst F1-scores of 0.43, 0.35, 0.30 and 0.28 respectively. Since HTML and CSS share a similar syntax and PLI was inept in classifying these languages, the F1-score for these languages dropped down.       

\begin{table}[!htb]
\centering
\begin{tabular}{|l|l|l|l|}
\hline \rowcolor{Gray}
\textbf{Programming} & \textbf{Precision} & \textbf{Recall} & \textbf{F1-score} \\ \hline
C           &  0.71   &   0.87     & 0.78     \\ \hline \rowcolor{Gray}
C\#         & 0.89  &    0.87    &  0.88     \\ \hline
C++         & 0.83   &   0.69    &  0.75      \\ \hline \rowcolor{Gray}
\end{tabular}
\caption{The performance of SCC for C, C\# and C++}
\label{family:pre}
\end{table}

\begin{table}[]
\centering

\begin{tabular}{|l|l|l|l|l|l|l|}
\hline\rowcolor{Gray}
\textbf{Performance}      & \multicolumn{2}{l|}{\textbf{Precision}} & \multicolumn{2}{l|}{\textbf{Recall}} & \multicolumn{2}{l|}{\textbf{F1-score}} \\ \hline
\textbf{Tools} & \textbf{SCC}       & \textbf{PLI}       & \textbf{SCC}      & \textbf{PLI}     & \textbf{SCC}       & \textbf{PLI}      \\ \hline\rowcolor{Gray}
Haskell        & \textbf{0.91}      & 0.58               & \textbf{0.86}     & 0.79             & \textbf{0.89}      & 0.67              \\ \hline
Python         & \textbf{0.91}      & 0.60                & \textbf{0.86}     & 0.69             & \textbf{0.88}      & 0.69              \\ \hline\rowcolor{Gray}
CSS            & \textbf{0.91}      & 0.76               & \textbf{0.82}     & 0.19             & \textbf{0.86}      & 0.30               \\ \hline
Lua            & \textbf{0.86}      & 0.37               & \textbf{0.81}     & 0.77             & \textbf{0.84}      & 0.50               \\ \hline\rowcolor{Gray}
Swift          & \textbf{0.87}      & 0.59               & \textbf{0.81}     & 0.49             & \textbf{0.84}      & 0.54              \\ \hline
Vb.Net         & 0.88               & \textbf{0.89}      & \textbf{0.79}     & 0.45             & \textbf{0.83}      & 0.60               \\ \hline\rowcolor{Gray}
C\#            & \textbf{0.76}      & 0.48               & \textbf{0.81}     & 0.54             & \textbf{0.79}      & 0.51              \\ \hline
JavaScript     & \textbf{0.77}      & 0.48               & \textbf{0.79}     & 0.48             & \textbf{0.78}      & 0.48              \\ \hline\rowcolor{Gray}
R              & 0.74               & \textbf{0.88}      & \textbf{0.79}     & 0.61             & \textbf{0.77}      & 0.72              \\ \hline
Markdown       & \textbf{0.73}      & 0.38               & \textbf{0.79}     & 0.22             & \textbf{0.76}      & 0.28              \\ \hline\rowcolor{Gray}
C              & \textbf{0.75}      & 0.58               & \textbf{0.76}     & 0.55             & \textbf{0.76}      & 0.56              \\ \hline
Bash           & 0.77               & \textbf{0.79}      & \textbf{0.75}     & 0.59             & \textbf{0.76}      & 0.67              \\ \hline\rowcolor{Gray}
Scala          & \textbf{0.80}       & 0.68               & 0.73              & \textbf{0.77}    & \textbf{0.76}      & 0.72              \\ \hline
PHP            & \textbf{0.79}      & 0.71               & \textbf{0.69}     & 0.55             & \textbf{0.74}      & 0.62              \\ \hline\rowcolor{Gray}
Perl           & \textbf{0.76}      & \textbf{0.76}      & \textbf{0.72}     & 0.63             & \textbf{0.74}      & 0.69              \\ \hline
Java           & \textbf{0.67}      & 0.57               & \textbf{0.73}     & 0.38             & \textbf{0.70}       & 0.46              \\ \hline\rowcolor{Gray}
Ruby           & \textbf{0.73}      & 0.29               & 0.67              & \textbf{0.79}    & \textbf{0.70}       & 0.43             \\ \hline
SQL            & 0.66               & \textbf{0.69}      & \textbf{0.63}     & 0.39             & \textbf{0.65}      & 0.50               \\ \hline\rowcolor{Gray}
Objective-c    & 0.48               & \textbf{0.85}      & \textbf{0.71}     & \textbf{0.71}    & 0.57               & \textbf{0.77}     \\ \hline
HTML           & \textbf{0.53}      & 0.33               & \textbf{0.56}     & 0.37             & \textbf{0.54}      & 0.35              \\ \hline\rowcolor{Gray}
C++            & \textbf{0.63}      & 0.62               & 0.44              & \textbf{0.69}    & 0.51               & \textbf{0.65}     \\ \hline
\end{tabular}
\caption{The comparison of SCC with PLI}
\label{my-label}
\end{table}

\begin{table}[!htb]
\centering
\begin{tabular}{|l|l|l|l|}
\hline \rowcolor{Gray}
\textbf{Programming} & \textbf{Precision} & \textbf{Recall} & \textbf{F1-score} \\ \hline
C\#-3.0     &     0.79   &   0.75   &   0.77     \\ \hline \rowcolor{Gray}
C\#-4.0      &   0.57    &  0.55    &  0.56      \\ \hline
C\#-5.0     &    0.56   &   0.59   &   0.58     \\ \hline \rowcolor{Gray}
\end{tabular}
\caption{The performance of SCC across various of C\#}
\label{cvarious}
\end{table}



\begin{table}[htbp]
\centering
\begin{tabularx}{\linewidth}{|l|X|}
\hline \rowcolor{Gray}
\multicolumn{1}{|c|}{\textbf{Programming Language }} & \multicolumn{1}{c|}{\textbf{Top 10 Features}}                                                   \tabularnewline \hline
C\# & add, asp, at, bool, byte, class, console, data, else, false
\tabularnewline \hline \rowcolor{Gray}
C++ & and, bool, boost, char, class, const, cout, cpp, data, double
\tabularnewline \hline  
C & and, argc, argv, array, break, buffer, case, char, const, count
\tabularnewline \hline
\end{tabularx}
\caption{The top $10$ features for C\#, C++ and C.}
\label{table:features1}   
\end{table}

\subsection{Discussion}
To summarize our results, the task of identifying the programming language of a code snippet seems to be fundamentally
different in nature compared to that of a source code file. While PLI is claimed to have an accuracy of 99.4\% for the
case of source code files, SCC, our tool based on a simple ML algorithm MNB, outperforms PLI for identifying the
programming language of code snippets from Stack Overflow. In Table \ref{my-label}, the comparison shows that SCC
achieves a higher F1 score for all programming languages except for Objective-C and C++. 
PLI particularly achieves a poor F1 score compared to SCC for CSS and HTML.
Objective-C has an equal
recall in both tools; however, the precision for PLI is much higher than SCC. This is because many code snippets are
misclassified by SCC as Objective-C. PLI also achieves a slightly higher precision than SCC for Vb.net, R, Bash and SQL.

\section{Threats to validity}
Internal validity: We studied how SCC and PLI are able to classify the programming language of a code
snippet. While we studied if SCC can distinguish between a family of programming languages (C, C\# and C++) and also
examined if SCC can identify the three versions of programming language C\#, we were unable to run those experiments for
PLI as it is closed-source. Also, we were only able to evaluate PLI with 150 snippets because we had to pay for every
classified snippet.

External validity: We only used Stack Overflow as the source of data for our analysis. We have not explored other sources such as GitHub repositories or other sources of code such as extracting a snippet from source code file of GitHub. Therefore, we cannot be absolutely confident that our results would be the same across all the sources of code snippets on programming languages. Our comparison is only against PLI which is the only tool available at this time. This is mainly due to the lack of open source tools for predicting languages.


\section{Conclusions}
In this paper we discussed the importance of predicting languages from code snippets. We argued that predicting the programming language from code snippets is far more challenging than from source code files considering complexity of today’s programming languages. We proposed Source Code Classification (SCC), a tool built using a Multinomial Naive Bayes classifier trained on a Stack Overflow dataset. SCC achieved an accuracy of 75\% and the average score for precision, recall and the F1 score with the proposed tool were 0.76, 0.75 and 0.75, respectively. We compared SCC against PLI, the only known proprietary tool for this problem and found that SCC achieved a much higher accuracy than PLI (that achieved only an accuracy of 55.5\%) on code snippets from Stack Overflow posts. 

\section*{Acknowledgments}
Kamel Alrashedy acknowledges the financial support of the Saudi Ministry of Education through a graduate scholarship. The authors would like to thank Dr Kwang Moo Yi for his helpful advice on this paper.

\end{document}